# The Secondary Star of the HgMn Binary φ Her

M. M. Dworetsky & Rosemary Willatt, Department of Physics & Astronomy
University College London
London WC1E 6BT, UK mmd@star.ucl.ac.uk

**Summary.** The binary nature of the bright ($V$ = 4.2 mag) Mercury-Manganese star φ Her has been known since 1976 and it was considered a low-amplitude single-lined SB. In a recent study we found evidence for lines of the secondary star. Other recent results from interferometry provide a good measure of the light ratio. It is very plausible that the secondary is a late main-sequence A star. We find the rotational velocity of the secondary to be ~42±5 km s$^{-1}$.

**Introduction.** The bright HgMn star φ Her was considered to be a single star until Aikman's (1976) discovery of low-amplitude orbital motion, with $K$ = 2.39 km s$^{-1}$ and a period of 560.5 day. Even after Aikman's discovery no sign of the secondary was found, up to and including the abundance analysis by Adelman et al. (1996). It was concluded that the light ratio of the pair must be large.

During a recent study of the spectrum, one of us (RW) found unusual depressions in the wings of some strong lines. We soon realised that we were seeing the spectrum of the faint secondary star. A synthesis analysis was undertaken with the spectrum synthesis package UCLSYN, which allows binary stars to be synthesised explicitly.

**Observations.** The spectra used were obtained at Lick Observatory with the Hamilton Échelle Spectrograph during the period 1995-2005. The resolution $R$ = 46,500 and the typical S/N is about 200. Two examples are shown in Figs. 1through 4. The Mg I line shows strong, broad wing extensions consistent with a relatively rapidly rotating stellar companion, while the Fe II line has only weak but definite wings also consistent with a cooler binary companion.

**Results.** The light ratio at $V$ is 2.64 mag according to the AAS abstract of Zavala et al (2004). These authors estimate the companion to be type A9 if it is a main sequence star, as seems likely. We adopted $T_{eff}$ = 11,650 K and log g = 4.00 for the primary, as in our previous work (e.g., Dworetsky and Budaj 2000). For the secondary, we adopted an assumed $T_{eff}$ = 7750 K, consistent with the suggested spectral type, and log $g$ = 4.3, consistent with a main sequence star. The effective temperatures and bolometric corrections of Flower (1996) were used to confirm that these values were consistent with the luminosity ratio of Zavala et al: log $L_B/L_A$ = -1.284 (bolometric), and we assumed masses of 3.1 and 1.7 solar masses for A and B respectively, based on Popper's (1980) masses for stars similar in $T_{eff}$ and luminosity to φ Her A and B.

For the purpose of spectrum synthesis of the secondary star we assumed the microturbulent velocity ξ = 3.5 km s$^{-1}$ in accord with Fig. 1 in Coupry & Burkhart (1992) for stars with similar effective temperatures.

The best fit rotational velocity for the secondary is 42 km s$^{-1}$ with an estimated uncertainty of ±5. For the primary, we adopted the value found by Jomaron et al (1999), 10.1 km s$^{-1}$, and ξ = 0.4 km s$^{-1}$ from Adelman (1988).

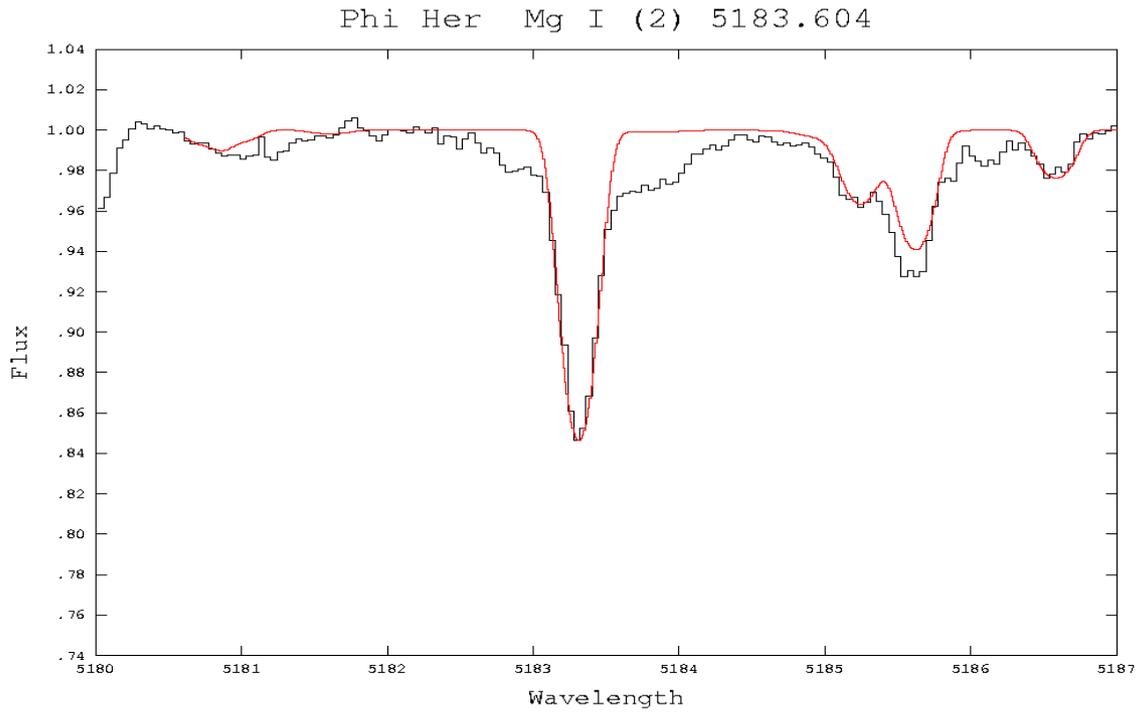

Figure 1. The Mg I line at 5183Å is shown together with a single-star synthesis (red) that beautifully fits the core of the line profile, but not the wings.

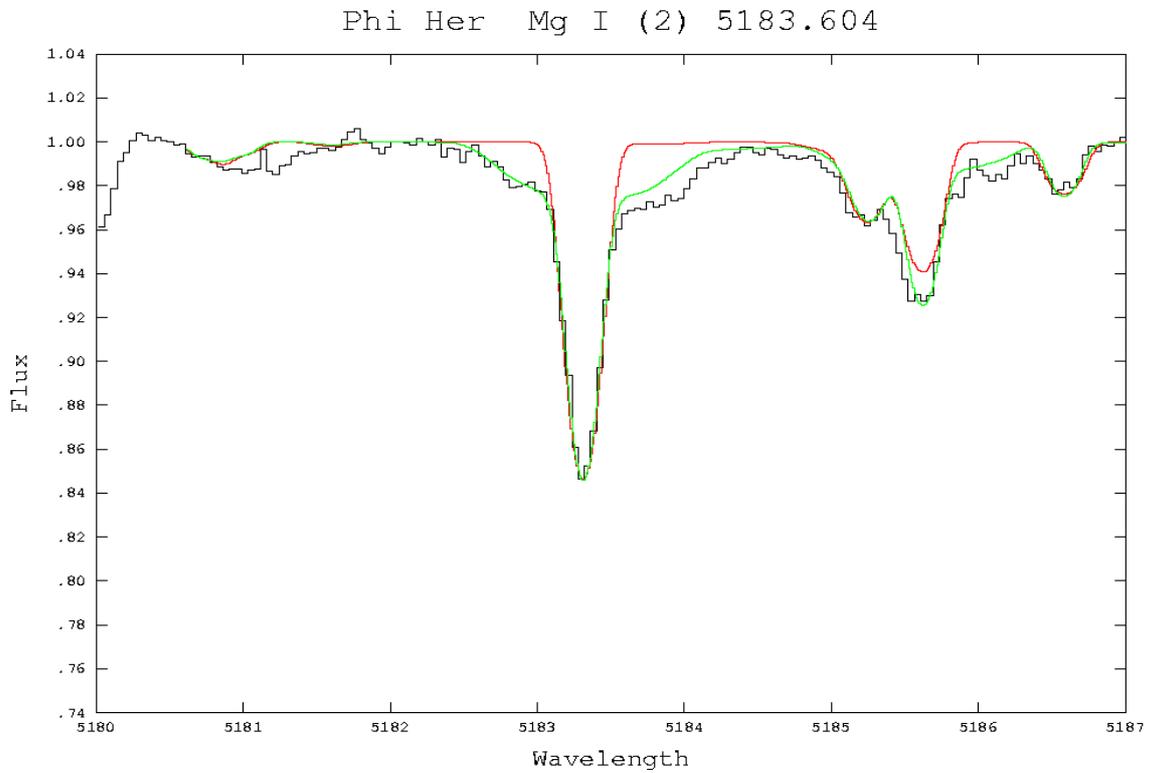

Figure 2. As Fig. 1, but with the addition of a secondary star (green) with $v\sin i$ = 42 km s$^{-1}$ and consistent with the light ratio discussed in the text. The fit isn't perfect but clearly there is a secondary star present.

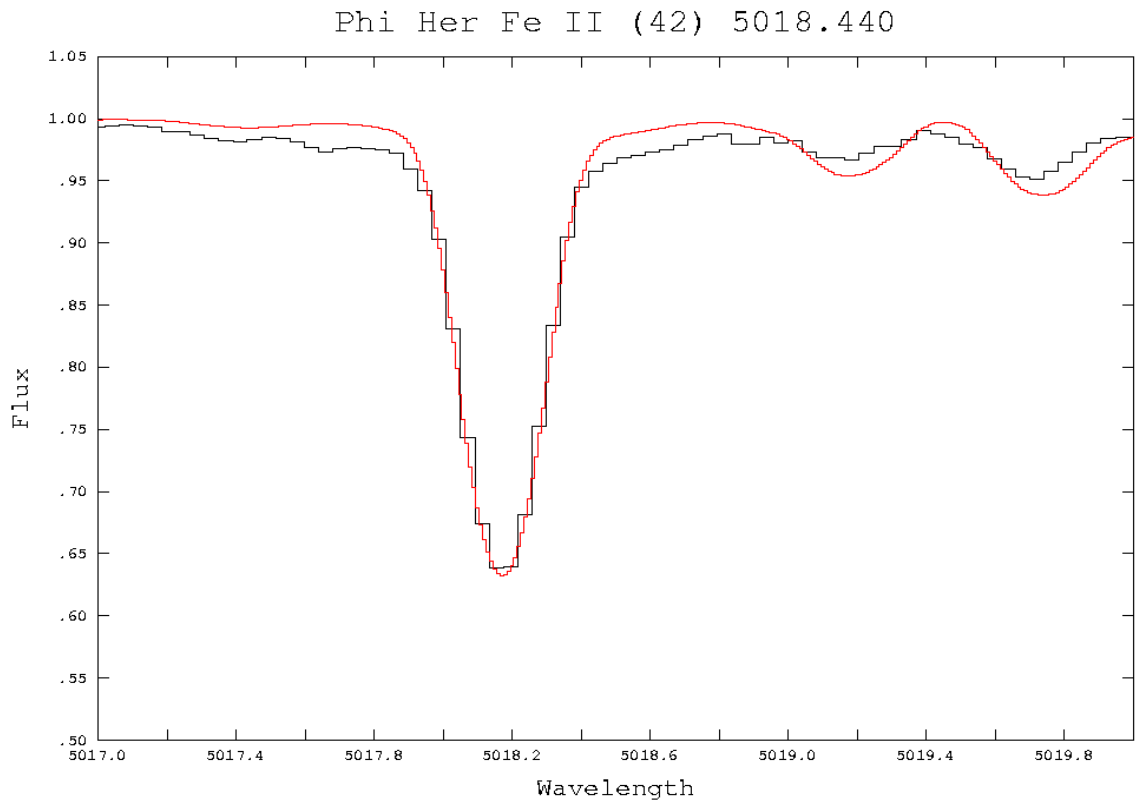

Figure 3. The Fe II line at 5018Å. Observations are in black. The synthesis assuming the *primary only*, with a best fit for the line near the centre, is shown in red. There appear to be wings for which the synthesis cannot account.

**Discussion.** This star now joins a considerable list of DLSB (double-lined spectroscopic binary) stars among the HgMn group. Without listing references in this poster, those we can mention include HR 4072, χ Lup, 46 Dra, 74 Aqr, κ Cnc, ι CrB, and 112 Her. There is absolutely no doubt that the secondary is some sort of late A star, probably on the main sequence.

**Acknowledgments.** We thank The Nuffield Foundation for its support for an Undergraduate Research Bursary for Rosemary Willatt, award number URB/02083. Observations obtained at Lick Observatory were supported by generous allotments of Guest Observer time by the Director, J. Miller, and financial support for travel was provided by the UK Particle Physics and Astronomy Research Council through its PATT grant to UCL.

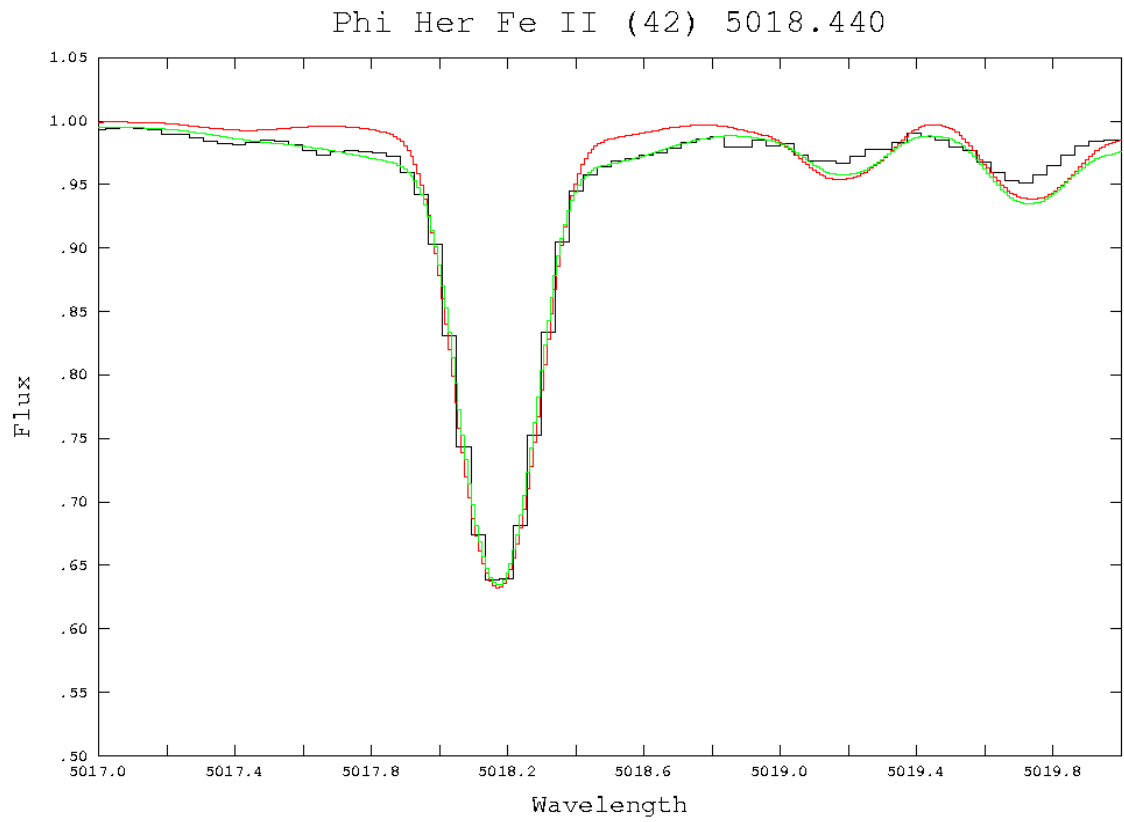

Figure 4. As Fig. 3, but with the synthesis of the secondary added in green. This is a good representation of the observed spectrum.